\documentclass[10pt,twocolumn,aps,prx,superscriptaddress,longbibliography,eprint]{revtex4-2}

\usepackage{graphicx}
\usepackage{dcolumn}
\usepackage{bm}
\usepackage[normalem]{ulem}
\usepackage[english]{babel}
\usepackage[T1]{fontenc}
\usepackage{amsmath}
\usepackage{amssymb}
\usepackage{float} 
\usepackage{braket}   
\usepackage{enumitem}
\usepackage[dvipsnames]{xcolor}
\usepackage[colorlinks=true, urlcolor=blue, linkcolor=blue,citecolor=blue, hypertexnames=false, unicode=true]{hyperref}
\usepackage{comment}
\usepackage{xcolor}
\usepackage{siunitx} 

\usepackage{cleveref}
\crefname{equation}{Eq.}{Eqs.}
\Crefname{equation}{Equation}{Equations}
\crefname{figure}{Fig.}{Figs.}
\Crefname{figure}{Figure}{Figures}
\crefname{section}{Sec.}{Sects.}
\Crefname{section}{Section}{Sections}
\crefname{table}{Table}{Tables}
\crefname{appendix}{Appendix}{Apps.}
\Crefname{appendix}{Appendix}{Apps.}

\begin{document}

\title{Exceeding the Parametric Drive Strength Threshold in Nonlinear Circuits}

\author{Mingkang Xia}
\altaffiliation{These authors contributed equally to this work.}
\affiliation{Department of Physics and Astronomy, University of Pittsburgh, Pittsburgh, PA 15213, USA}
\affiliation{Department of Applied Physics, Yale University, New Haven, CT 06511, USA}
\author{Crist{\'o}bal Lled{\'o}}
\altaffiliation{These authors contributed equally to this work.}
\affiliation{Institut Quantique and Département de Physique, Université de Sherbrooke, Sherbrooke J1K 2R1 QC, Canada}
\affiliation{Departamento de Física, Facultad de Ciencias Físicas y Matemáticas, Universidad de Chile, Santiago 837.0415, Chile}
\author{Matthew Capocci}
\affiliation{Department of Physics and Astronomy, Northwestern University, Evanston, Illinois 60208, USA}
\author{Jacob Repicky}
\affiliation{Department of Applied Physics, Yale University, New Haven, CT 06511, USA}
\author{Benjamin D'Anjou}
\affiliation{Institut Quantique and Département de Physique, Université de Sherbrooke, Sherbrooke J1K 2R1 QC, Canada}
\author{Ian Mondragon-Shem}
\affiliation{Department of Physics and Astronomy, Northwestern University, Evanston, Illinois 60208, USA}
\author{Ryan Kaufman}
\affiliation{Department of Physics and Astronomy, University of Pittsburgh, Pittsburgh, PA 15213, USA}
\author{Jens Koch}
\affiliation{Department of Physics and Astronomy, Northwestern University, Evanston, Illinois 60208, USA}
\author{Alexandre Blais}
\affiliation{Institut Quantique and Département de Physique, Université de Sherbrooke, Sherbrooke J1K 2R1 QC, Canada}
\affiliation{Canadian Institute for Advanced Research, Toronto, M5G 1M1 Ontario, Canada}
\author{Michael Hatridge}
\affiliation{Department of Applied Physics, Yale University, New Haven, CT 06511, USA}

\date{\today}

\begin{abstract}
   Superconducting quantum circuits rely on strong drives to implement fast gates, high-fidelity readout, and state stabilization. However, these drives can induce uncontrolled excitations---so-called ``ionization''---that compromise the fidelity of these operations.
   While now well-characterized in the context of qubit readout, it remains unclear how general this limitation is across the more general setting of parametric control. Here, we demonstrate that a nonlinear coupler, exemplified by a transmon, undergoes ionization under strong parametric driving, leading to a breakdown of coherent control and thereby limiting the accessible gate speeds. Through experiments and numerical simulations, we associate this behavior with the emergence of drive-induced chaotic dynamics, which we characterize quantitatively using the instantaneous Floquet spectrum. Our results reveal that the Floquet spectrum provides a unifying framework for understanding strong-drive limitations across a wide range of operations on superconducting quantum circuits. This insight establishes fundamental constraints on parametric control and offers design principles for mitigating drive-induced decoherence in next-generation quantum processors.
\end{abstract}

\maketitle

\section{Introduction}

Superconducting circuits provide a leading platform for quantum information processing~\cite{Blais2021Circuit,Zhu2022Advantage,Acharya2024QEC}, yet the performance of these systems remains limited by loss mechanisms and undesired interactions, particularly under strong drive conditions. One striking example is the breakdown of quantum non-demolition readout in superconducting qubits, which has been linked to multi-photon resonances that excite the qubit outside of its computational manifold \cite{Jeffrey2014Fast,Sank2016Measurement,Minev2019,Khezri2023Measurement,Shillito2022Dynamics,Cohen2023Reminiscence,Dumas2024,Xiao2023Diagrammatic,Bista2025_FluxoniumIonization,Nesterov2024}.
This ``ionization'', where the qubit is unintentionally excited due to strong drive and nonlinearity, is now understood as a fundamental limit of measurement in circuit QED and is governed by the same phenomenology as the ionization of highly excited hydrogen atoms exposed to microwave fields~\cite{Breuer1989_Floquet_Ionization}.

The same combination of strong drive and nonlinearity that leads to qubit ionization also appears in a broader class of operations, and may therefore represent an intrinsic limitation to their performance. For example, qubit reset~\cite{Magnard2018_AllMicrowaveReset,Egger2018_PulsedReset,Zhou2021_ParametricResetTunableQubit}, quantum state stabilization~\cite{Grimm2020_KerrCatExperiment,Hajr2024_2DKerrCat,Frattini2024_Sqeezed_Kerr,Reglade2024_DissipativeCat}, as well as controllable interactions between pairs of qubits~\cite{Caldwell2018_Parametric_Gate,Yan2018Tunable, Reagor2018Parametric,Wu2021Strong,Zhao2022Tunable,Xia2024Subharmonic} and bosonic modes \cite{Gao2018,Gao2019,Chapman2023} are often achieved using parametric control, where a specific interaction is activated using a strong, off-resonant drive on a nonlinear system. Increasing the drive amplitude in these processes is expected to speed up the operations. However, this often leads to unexpected behavior, ultimately limiting the gate speed and efforts to achieve high-fidelity control. As superconducting processors advance toward scalable, fault-tolerant operation, understanding and mitigating these limitations becomes critical.

In this work, we investigate whether a nonlinear element---such as a qubit or qubit-based coupler---undergoes ionization during parametric driving and whether the underlying mechanism shares the same origin as in qubit readout. To this end, we study a transmon qubit driven at a frequency much lower than its $0 - 1$ transition, mimicking the conditions of a coupler driven at a qubit-qubit difference frequency. We observe that at a sufficiently strong drive, the transmon undergoes uncontrolled excitations into high-energy states, marking a clear breakdown of coherent control of parametric processes. We show that this behavior is well captured by numerical simulations that fully account for the Josephson cosine potential and find remarkable agreement with an analysis based on the instantaneous Floquet spectrum. Moreover, beyond the threshold drive amplitude for ionization, the occupation probability distribution of the transmon state is very sensitive to the drive amplitude, a behavior that is reminiscent of classical chaos in the transmon~\cite{Cohen2023Reminiscence}.

Our results suggest that the breakdown of parametric control in driven superconducting circuits is a fundamental, generalizable phenomenon with implications for fast parametric gates, high-fidelity readout, and, more generally, any scenario involving strong drives  at low frequencies on a superconducting qubit or coupler. By identifying the onset of chaotic behavior as the key limiting factor using a simple Floquet analysis, we provide a theoretical framework that can inform the design of next-generation superconducting devices.

\section{Results and Discussion}

\begin{figure*}[t]
    \centering
    \includegraphics[width=2\columnwidth]{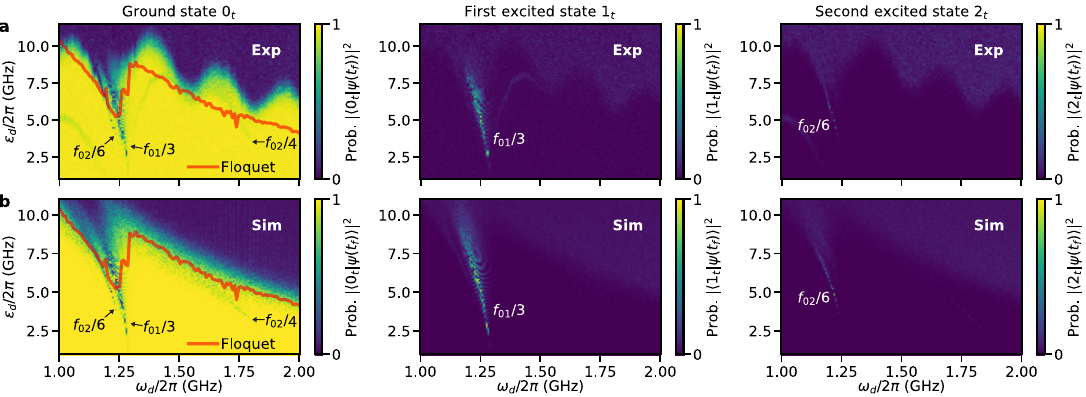}
    \caption{{\bf Strong drive limit to parametric processes.}
    {\bf a}, Measured occupation of the first three transmon levels after preparing the ground state and applying a $\SI{100}{ns}$ flat-top pulse of varying amplitude ($\varepsilon_d$) and frequency ($\omega_d$). {\bf b}, Numerical simulation of the experiment averaged over 10 values of the gate charge $n_g$.
    Parametric processes are activated at specific frequencies, as indicated by the Rabi fringes. There is an amplitude threshold for all drive frequencies beyond which the transmon is excited away from the ground state into highly excited states. Two features appearing in the experimental data but absent in simulation correspond to transitions activated in the presence of thermal population of the resonator: the direct $\ket{0_t, 1_r}\to\ket{2_t, 0_r}$ and the two-photon $\ket{0_t, 1_r}\to\ket{1_t, 0_r}$ processes, where $\ket{n_r}$ is the resonator state with $n_r$ photons. The red line indicates the threshold amplitude for ionization as obtained from Floquet theory.
    }
    \label{fig:Figure_1}
\end{figure*}

We use a transmon qubit of frequency $\omega_{01}/2\pi =  
\SI{3876.7}{MHz}$ and negative anharmonicity $\alpha/2\pi=148.6$~MHz ($\hbar = 1$). For dispersive readout, the transmon is coupled with strength $g/2\pi = \SI{60.1}{MHz}$ to a $\lambda/2$-resonator with bare frequency $\omega_{r}/2\pi = \SI{6404.3}{MHz}$ (see Methods and Supplemental Fig.~S1 for details). The transmon is first initialized in its ground state and then driven with a tanh-edged square pulse of amplitude $\varepsilon_d$, total duration $t_\text{pulse} = \SI{100}{ns}$, and  $t_\text{ramp} = \SI{5}{ns}$ ramp-up and ramp-down times. To mimic the typical conditions of a parametric coupler, the drive frequency $\omega_d/2\pi \in [1,2]~\text{GHz}$ is chosen to be much lower than the qubit frequency. Consequently, the qubit-drive detuning $\Delta_{qd} = (\omega_{01} - \omega_d)$ is always positive. 
The resulting state occupation probabilities are measured using a dispersive readout capable of clearly resolving the first five transmon states.
In \cref{fig:Figure_1}a, we show the measured occupation in the ground, first-excited, and second-excited states, sweeping over the drive frequency and amplitude. We denote the $i$-th transmon state as $\ket{i_t}$ and define $\omega_{ij}$ as the transition frequency from $\ket{i_t}$ to $\ket{j_t}$.

Focusing first on the region of low drive amplitudes, we observe that the qubit remains in the ground state for most drive frequencies, consistent with the large drive-qubit detuning. We also observe distinct subharmonic resonances with Rabi fringes where multiple photons from the drive are upconverted into a smaller number of transmon excitations. The most prominent resonance occurs when the drive frequency equals $\omega_{01}/3=2\pi\times\SI{1292.3}{MHz}$, where three drive photons are converted into a single excitation through the four-wave mixing allowed by the transmon's nonlinearity. Another strong resonance appears at $\omega_{02}/6=2\pi\times\SI{1267.5}{MHz}$, which is at a frequency $\alpha/6$ below the previous feature. There, six photons from the drive are upconverted into two transmon excitations via eight-wave mixing. These processes are well-understood within the Kerr approximation of the transmon, and can be exploited to realize fast and high-fidelity gates~\cite{Xia2024Subharmonic}. As discussed in more detail below, these resonances allow us to calibrate the drive amplitude. Additional weaker resonances involving the qubit and the readout resonator are also observed, see figure caption.

\begin{figure}[t]
    \centering
    \includegraphics[width=1\columnwidth]{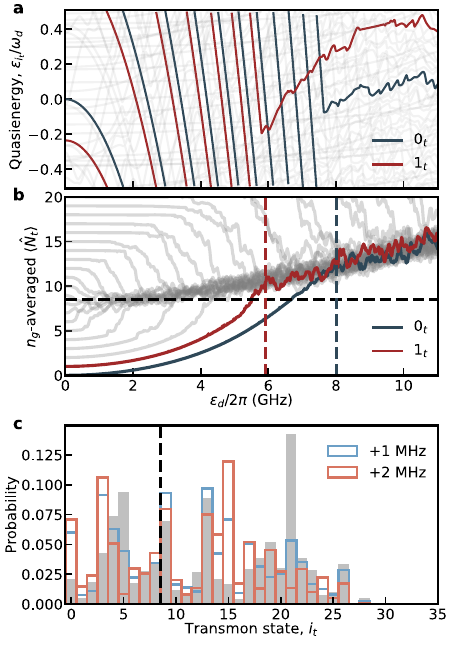}
    \caption{{\bf Chaotic behavior in the Floquet spectrum and parameter sensitivity.}
    {\bf a}, Quasienergy spectrum as a function of the drive amplitude for a fixed drive frequency $\omega_d/2\pi = \SI{1.4}{GHz}$ and gate charge $n_g=0$.
    {\bf b}, Average population of Floquet modes as a function of drive amplitude for the same drive frequency, averaged over the gate charge.
    The horizontal dashed line indicates the number of energy levels confined within the transmon’s cosine potential.
    The blue and red vertical dashed lines indicate the computed ionization threshold for the ground- and first-excited states, respectively.
    {\bf c}, Transmon occupation distribution at the end of the pulse for a drive with the same frequency and amplitude of \(\SI{10}{GHz}\) and a single value of gate charge ($n_g=0$). Only the first nine states are within the cosine potential (vertical dashed line). Additional distributions for drive amplitudes differing by \(\SI{1}{MHz}\) illustrate sensitivity to drive parameters.
    }
    \label{fig:Figure_2}
\end{figure}

Focusing now on the region of high drive amplitudes, \cref{fig:Figure_1}a, we observe a transition at a threshold drive amplitude beyond which the transmon is excited out of the ground state at all drive frequencies. This feature limits the speed of the subharmonic gates and, more generally, of transmon-based qubit couplers. Beyond this threshold, we observe that the qubit population does not accumulate in the first- or second-excited states but instead spreads into many energy levels, see \cref{fig:Figure_2}b,c and Supplementary Figs.~S2 and S3. We note that the oscillatory pattern in the experimentally observed amplitude threshold is attributed to uncalibrated, frequency-dependent transmission due to impedance mismatch causing resonances in the short cables between the bottom of the mixing chamber and the sample housing. 

While the subharmonic resonances are captured by the Kerr approximation of the transmon Hamiltonian, the observed amplitude threshold is not. For this reason, we turn to exact numerical simulations of the time dynamics of the transmon without expansion of the cosine potential. Moreover, because high-energy states of the transmon are involved, we use a model which accounts for the transmon's offset charge~\cite{Dumas2024,Cohen2023Reminiscence,Khezri2023Measurement,Fechant2025} and higher-order harmonics of the cosine potential~\cite{Willsch2024_HigherTransmonHarmonics,Fechant2025,Wang2025}. The system is described by the Hamiltonian \begin{equation} \label{eq: driven transmon Hamiltonian}
\begin{split}
    \hat H_t(t) =& 4E_C (\hat n_t-n_g)^2 - \sum_{m=1}^4 E_{J,m} \cos(m\hat \varphi_t) \\
    &+ \varepsilon_d(t) \sin(\omega_d t)\hat n_t.
\end{split}
\end{equation}
Here, $\hat n_t$ and $\hat \varphi_t$ are the transmon's charge and phase operators, $E_C$ is the charging energy, $E_{J,m}$ are the energies of the different Josephson harmonics, and $n_g$ is the offset charge. The transmon parameters $E_C$ and $E_{J,m}$ are fitted to independently measured spectroscopy data, see Supplemental information Sec.~III. We also account for the presence of the readout resonator (see Methods).  Starting from the dressed ground state, we numerically integrate the Schr\"odinger equation under the full qubit-resonator Hamiltonian. From this, we evaluate the transmon occupation probabilities at the final time, following the application of the same $\SI{100}{ns}$-long pulse used in the experiment (see Methods). \Cref{fig:Figure_1}b shows the resulting occupation for the transmon's first three states. The agreement with the experimental data is excellent, capturing both the subharmonic resonances and the drive amplitude threshold. This level of agreement allows us to calibrate the conversion factor between room-temperature voltage and transmon drive amplitude using the Rabi oscillations of the $\omega_{01}/3$ resonance (see Methods). Surprisingly, the numerical simulations reveal that above the threshold---e.g., at $\varepsilon_d/2\pi = \SI{10}{GHz}$---the transmon occupation can spread to as many as thirty levels (see \cref{fig:Figure_2}c). Given that only nine levels lie within the cosine potential of this transmon, the strongly detuned drive promotes significant occupation of states above the top of the cosine well, where the eigenstates become highly sensitive to gate charge fluctuations. In the context of measurement, this phenomenon has been referred to as measurement-induced state transition~\cite{Sank2016Measurement,Khezri2023Measurement} and transmon ionization~\cite{Dumas2024,Shillito2022Dynamics,Cohen2023Reminiscence}. Our results clearly show that this phenomenology is not limited to qubit readout.

This transition from the qubit state to a large number of excited states can be understood as a result of the negative transmon anharmonicity $\alpha$ combined with the positive qubit-drive detuning $\Delta_{qd}$~\cite{Cohen2023Reminiscence,Khezri2023Measurement,Dumas2024}. Indeed, while a low-amplitude off-resonant drive acts as a small perturbation on the computational states, transitions $\omega_{i_t, i_t+1}$ become increasingly resonant with the drive when considering higher states of the cosine potential. Because of the small anharmonicity, there is typically a transition for which $|\omega_{i_t^\star,i_t^\star+1} - \omega_d| \sim \varepsilon_d$ where the dispersive qubit-drive approximation fails. Accounting for the transmon’s charge dispersion, here this occurs for $i_t^\star \sim 9_t$. For that state, the drive does not act as a perturbation and induces a strong hybridization with a neighbor state, even at small amplitudes. As the drive strength increases, this drive-induced hybridization spreads to neighboring states and eventually reaches the computational subspace. At that point, strong hybridization is expected across much of the transmon spectrum resulting in the observed threshold. Additional consequences of this mechanism include an increase in the ionization threshold with qubit-drive detuning and a lower threshold when the transmon is initialized in $\ket{1_t}$ or $\ket{2_t}$ (see \cref{fig:Figure_3} and Fig.~S3 of the Supplemental information).

A similar phenomenology leading to strong hybridization arises in dispersive qubit readout at positive qubit-resonator detuning~\cite{Cohen2023Reminiscence,Khezri2023Measurement,Dumas2024}. In that context, the smaller detuning---used to enhance the dispersive shift---leads to additional multiphoton resonances that precipitate ionization at specific frequencies~\cite{Khezri2023Measurement}. In contrast, for the large detuning considered here, no single resonance dominates when gate charge fluctuations are considered, resulting in a broadly monotonic dependence of the ionization threshold on drive frequency away from the identified $\omega_{01}/3$ feature, see Supplemental information Sec.~IV.

This phenomenology can be understood more clearly using Floquet theory for periodically driven quantum systems, which can be generalized to account for situations in which the Hamiltonian is not strictly periodic in time, such as here where the drive amplitude is varied~\cite{Breuer1989QuantumPhases}. This method has been used to study the ionization of highly excited hydrogen atoms under microwave drive~\cite{Breuer1989_Floquet_Ionization} and to accurately predict transmon ionization during dispersive readout~\cite{Dumas2024}. In this approach, we neglect the presence of the measurement resonator and the time dependence of the drive amplitude. We thus use the Hamiltonian of \cref{eq: driven transmon Hamiltonian} with $\varepsilon_d(t) \rightarrow \varepsilon_d$, making the system time-periodic with period $T = 2\pi/\omega_d$. The dynamics are then governed by the Floquet modes $\ket{\phi_{i_t}(t)}$ and quasienergies $\epsilon_{i_t}$, which satisfy the eigenvalue equation $\hat{U}(t+T,t)\ket{\phi_{i_t}(t)} = e^{-i\epsilon_{i_t} T} \ket{\phi_{i_t}(t)}$~\cite{Grifoni1998}. These quantities are obtained by numerically diagonalizing $\hat{U}(t+T,t)$, the unitary propagator over one period of the drive. Because the quasienergies are extracted from a phase, they are only defined modulo the drive frequency $\omega_d$. At zero drive amplitude, the quasienergies correspond to the transmon energies modulo $\omega_d$, and the Floquet modes $\ket{\phi_{i_t}(t)}$ reduce to the bare transmon eigenstates $\ket{i_t}$. From this starting point, the quasienergies and modes can be unambiguously labeled by a unique transmon index $i_t$ at every drive amplitude~\cite{Dumas2024}. 

Crucially, an avoided crossing between two quasienergies indicates a multiphoton resonance: two transmon energy levels, shifted by the drive, would become degenerate modulo $\omega_d$, but this degeneracy is lifted by the drive field, which couples the levels via the exchange of an integer number of photons at $\omega_d$.
\Cref{fig:Figure_2}a shows the quasienergy spectrum as a function of drive amplitude for the first 30 transmon states. The quasienergies corresponding to the states $0_t$ and $1_t$ are highlighted in blue and red, respectively, and exhibit two distinct types of behavior. First, at low drive amplitudes, the quasienergies are ac-Stark shifted by the off-resonant drive; the apparent discontinuities are due to the modular nature of these quantities. Second, above a qubit-state-dependent amplitude threshold, the quasienergies display numerous kinks, each corresponding to a resonance. As a result, during time evolution starting from the bare state $\ket{0_t}$, the system follows the instantaneous Floquet mode associated with the ground state, $\ket{\phi_{0_t}}$, until it encounters a resonance and hybridizes with another mode~\cite{Dumas2024}. In contrast to dispersive readout, where one or a few resonances dominate, the large positive detuning considered here leads to multiple relevant resonances and strong hybridization across many states.

The strong hybridization is even more clearly seen by considering the average population of the Floquet modes, $\bra{\phi_{i_t}(0)} \hat N_t \ket{\phi_{i_t}(0)}$ with $\hat N_t = \sum_{i_t} i_t \ket{i_t} \bra{i_t}$ the transmon number operator, as a function of the drive amplitude for a fixed drive frequency, see \cref{fig:Figure_2}b. Because hybridization involves highly excited states of the transmon, the results are averaged over gate charge~\cite{Cohen2023Reminiscence,Dumas2024,Fechant2025,Wang2025}. At low amplitudes, the average population in modes $\ket{\phi_{0_t}}$ (blue) and $\ket{\phi_{1_t}}$ (red) increase gradually, consistent with the dispersive hybridization of transmon states under an off-resonant drive. At higher amplitudes, however, these modes strongly hybridize with high-energy modes, as indicated by a sharper increase in population followed by a bunching of the different populations \cite{Dumas2024}. As expected from the discussion above, this bunching emerges at very low drive amplitude at $i_t^\star = 9_t$. As the drive amplitude increases, the qubit-drive dispersive approximation fails for neighboring states, which then join the bunching layer pair by pair. In \cref{fig:Figure_2}b, the blue (red) dashed line indicates the drive amplitude at which the ground-state (excited-state) Floquet mode enters the bunching layer (see Methods). This predictor is obtained by solving the time-dependent Schrödinger equation over a single drive period to extract the Floquet spectrum, which is more efficient than simulating the full pulse dynamics.
It shows excellent agreement with both the dynamical simulations and experimental observations (see \cref{fig:Figure_1}).
These results emphasize that the onset of ionization is here not attributable to a single accidental multiphoton resonance but rather to a persistent and strong hybridization involving many Floquet modes. While a cleverly designed subharmonic pulse might allow the state to diabatically follow the instantaneous Floquet mode $\ket{\phi_{0_t}}$ or $\ket{\phi_{1_t}}$ across an isolated resonance, the onset of strong hybridization results in a drive amplitude threshold that is exceptionally difficult to surpass without ionizing the qubit. 

The population clustering observed in \cref{fig:Figure_2}b has an interesting connection to classical chaos in the driven rotor or rigid pendulum,
the classical analog of the transmon~\cite{Koch2007ChargeInsensitive, Cohen2023Reminiscence}. In the classical case, a small amplitude drive creates a chaotic layer in phase space around the separatrix at energy $2 E_J$, resulting from the system's instability to small perturbations~\cite{Zaslavsky2005}. As the drive amplitude increases, this chaotic layer expands, eventually engulfing the region of phase space associated with small-amplitude oscillations of the pendulum. 
As discussed in Refs.~\cite{Cohen2023Reminiscence, Dumas2024}, ionization is expected to occur when the orbits in phase space corresponding to the qubit-states Floquet modes $\ket{\phi_{0_t,1_t}}$ join the chaotic layer, results that are in qualitative agreement with \cref{fig:Figure_1}, see Supplemental information Sec.~V. In the quantum case, population clustering at small drive amplitudes involves Floquet modes associated with transmon energy states around $2 E_J$, corresponding to the top of the transmon potential well and the separatrix in the classical model (horizontal dashed line in \cref{fig:Figure_2}b). With increasing drive amplitude, additional Floquet modes are drawn into the cluster, and eventually, the computational modes $0_t$ and $1_t$ join, too. This behavior of the quantum system mirrors that of the classical system's chaotic layer engulfing the smallest orbits in phase space. The sensitivity to initial conditions of the chaotic classical system manifests as strong parameter dependence in the quantum case~\cite{Peres1984}. Prior work has shown that the quasienergy spectrum exhibits level repulsion, with splittings following a Wigner-Dyson distribution when sampled over gate charge~\cite{Cohen2023Reminiscence}. 
As a complementary illustration of chaos-induced sensitivity, \cref{fig:Figure_2}c shows in gray bars the numerically obtained transmon occupation probability distribution after a 100\,ns pulse of frequency $\omega_d/2\pi = \SI{1.4}{GHz}$ and amplitude $\varepsilon_d/2\pi = \SI{10}{GHz}$ (above the threshold). The occupation extends well above the top of the cosine potential (vertical dashed line). Adjusting the drive amplitude by as little as 0.1\% (blue bars) or 0.2\% (orange bars) results in a substantial change in the final transmon occupation distribution.

\begin{figure}[t]
    \centering
    \includegraphics{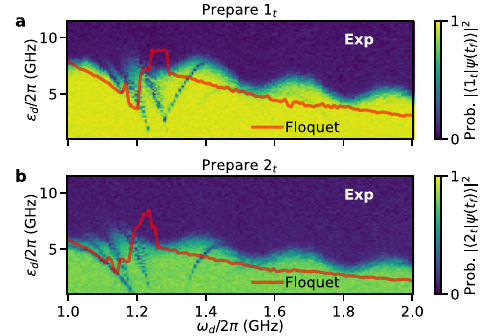}
    \caption{{\bf Lowering of threshold drive amplitude in higher transmon states.}
    {\bf a}, Measured $1_t$ state population after preparing the transmon in state $1_t$ and applying a \SI{100}{ns} flat-top pulse. The $\omega_{01}/3$, $\omega_{12}/3$, and $\omega_{13}/6$ resonances are prominent in the 1.2-1.3\,GHz region. An additional resonance (which shifts toward higher frequency) is also visible, which represents a two-photon $\ket{1_{t},0_{r}}\rightarrow\ket{0_t,1_r}$ transition. {\bf b}, Measured $2_t$ state population after preparing the transmon in state $2_t$ and applying a \SI{100}{ns} flat-top pulse. Here, the $\omega_{12}/3$, $\omega_{23}/3$, and $\omega_{24}/6$ resonances are prominent near 1.2\,GHz. When preparing $\ket{2_t}$, two resonator-involved transitions appear. The lower energy resonance (shifting toward lower frequency) is a direct $\ket{2_t,0_r}\rightarrow\ket{0_t,1_r}$ transition, and the higher energy resonance (shifting to higher frequency) is a two-photon $\ket{2_t,0_r}\rightarrow\ket{1_t,1_r}$ transition. In both figures, the red line indicates where the Floquet analysis predicts the entrance of the initial state into the band of strongly hybridized Floquet modes. The accuracy of the threshold prediction breaks down near the strong resonances due to hybridization with the $\ket{0_t}$ Floquet mode, which has a higher threshold amplitude.
    }
    \label{fig:Figure_3}
\end{figure}

Based on the results above, states closer to the top of the cosine potential well are expected to enter the band of strongly hybridized modes at lower drive amplitudes than the ground state. Thus, we expect to find a lower threshold drive amplitude when initializing the qubit in a higher energy state. To demonstrate this, we perform experiments in which the transmon is first prepared in $\ket{1_t}$ (\cref{fig:Figure_3}a) or $\ket{2_t}$ (\cref{fig:Figure_3}b) using resonant $\pi$-pulses, followed by the same protocol as in \cref{fig:Figure_1}a. Comparing the results of these experiments with predictions of the threshold condition from Floquet simulations (red line) yields overall excellent agreement. In both cases, the threshold drive amplitude is substantially reduced for a successively higher energy initial state, with $\varepsilon_{d,\mathrm{thresh}}$ reduced by nearly a factor of two for $\ket{2_t}$ compared to $\ket{0_t}$. 

\begin{figure}[t]
    \centering
    \includegraphics{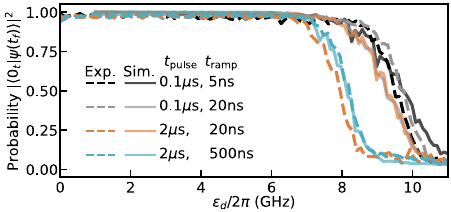}
    \caption{{\bf Dependence of the transmon breakdown threshold on pulse shape of the drive.
    } Experimental data (dashed) and numerical simulation (solid) showing the onset of breakdown using four pulse shapes differing in total pulse lengths and ramp times. The frequency of the drive is $\omega_d/2\pi=$\SI{1.3}{GHz} in a region free from low-amplitude subharmonic resonances such that the only change in $\ket{0_t}$ population is attributed to the transmon entering the breakdown regime. The simulation follows the same approach as \cref{fig:Figure_1} and also includes $n_g$ averaging.}
    \label{fig:Figure_4}
\end{figure}

We now investigate the effect of the pulse shape on the observed threshold. \Cref{fig:Figure_4} presents experimental (dashed) and numerical (solid) results for various ramp times and overall pulse durations at a drive frequency slightly above the $\omega_{01}/3$ resonance, a region where no low-amplitude subharmonic resonances of consequence are present. First, we observe that slower ramp times lead to ionization at lower drive amplitudes. This can be understood from the perspective of a Landau-Zener process. Indeed, during the ramp, the state adiabatically follows a Floquet mode until it encounters a resonance. At the avoided crossing, a diabatic passage leaves the character of the state unchanged, whereas an adiabatic passage allows hybridization with another mode. As a result, weaker resonances start inducing ionization if the ramp time is increased~\cite{Dumas2024,Wang2025,Shillito2022Dynamics}.
Second, the pulse duration has no effect in the simulations, as demonstrated by the two pulses with $t_\text{ramp} = \SI{20}{ns}$. Since dissipation is not included in the simulations, the pulse duration only affects the Landau-Zener-Stückelberg phase accumulated during the flat portion of the pulse. Averaging over gate charge effectively eliminates this phase dependence. In contrast, experimental data show a clear dependence on pulse duration. A possible explanation is that spontaneous decay and emission, which are not accounted for in the simulations, enable transitions to other Floquet modes that ionize earlier. Consequently, the longer pulse ($t_\text{pulse} = \SI{2}{\mu s}$, $t_\text{ramp} = \SI{20}{ns}$, orange) leads to ionization at lower drive amplitude compared to the shorter one ($t_\text{pulse} = \SI{0.1}{\mu s}$, $t_\text{ramp} = \SI{20}{ns}$, gray).

\section{Conclusion}

Through experiments and simulations, we have shown that when driven too strongly off-resonance, the transmon's computational states invariably undergo transitions into a broad distribution of highly excited states and must be reset before coherent operations can resume. 
We identify the evolution in this regime as an irreversible state transition owing to the sensitivity to parameters of the strongly hybridized Floquet modes~\cite{Peres1984, Cohen2023Reminiscence}.  Importantly, the resulting occupation spread across many energy levels suggests that fast and reliable reset is extremely challenging in this regime.

We expect this threshold behavior to be a generic feature of low-impedance Josephson-junction-based circuits, potentially accounting for previously reported reductions in fidelity during high-power parametric operations~\cite{Chapman2023}. In the transmon, this chaotic-like behavior begins in states near the top of the cosine potential well at low drive frequencies and progressively spreads throughout the spectrum. However, neither the shape of the potential nor the presence of such initial localized hybridization is essential for the emergence of strong hybridization among Floquet modes, suggesting the phenomenon may arise more broadly in other circuits.

Based on these findings, we hypothesize that driving a single mode beyond its threshold could strongly perturb other coupled modes in a multi-qubit device, potentially leading to cascaded instabilities requiring a large amount of resources to correct. Therefore, quantum computing architectures relying on parametric control schemes should benefit from the results presented here. Moving forward, we aim to explore similar regimes in larger systems where two-qubit gates and devices with substantially different design parameters can be explored.

\section{Methods}

\subsection{Time dynamics simulations}

The results of \cref{fig:Figure_1}b) are obtained by solving the Schrödinger equation with the time-dependent Hamiltonian
\begin{equation} \label{eq:full transmon-resonator Hamiltonian}
\begin{split}
\hat{H}_{tr}(t) = \omega_r \hat a^\dag \hat a + \hat{H}_t(t)
+ ig (\hat n - n_g)(\hat a^\dag - \hat a),
\end{split}
\end{equation}
where the transmon Hamiltonian [c.f.~\cref{eq: driven transmon Hamiltonian}] includes contributions from higher-order Josephson harmonics with energies $E_{J,m}$ and is dispersively coupled to the readout resonator (with annihilation operator $\hat a$).

We initialize the system in the dressed ground state $\ket{\overline{0_t, 0_r}}$, where the first and second labels correspond to the transmon and resonator excitations, respectively. We then drive the transmon with fixed frequency $\omega_d$ and time-dependent amplitude $\varepsilon_d(t) = \varepsilon_d \lambda(t)$, using a tanh-box pulse-edge given by
\begin{equation}
\begin{split}
\lambda(t) = \frac{1}{1-c_1} \bigg[ &\frac{\tanh(kt+c_0)}{2} - \frac{\tanh[k(t-t_f)]}{2} \\
& - \frac{c_0}{2} - c_1\bigg],  
\end{split}
\end{equation}
with $c_0=\arctan(2c_1-1)$, $c_1=0.01$, and $k=(\arctan(2\times 0.95-1) - c_0)/t_\text{ramp}$, pulse duration $t_f=\SI{100}{ns}$, and ramp-up and ramp-down time $t_\text{ramp} = \SI{5}{ns}$.

To obtain the ground state probability in the first panel of \cref{fig:Figure_1}b, we compute the sum of the probabilities of finding the state in any of the dressed states $\ket{\overline{0_t, n_r}}$ at time $t_f$, for all photon number $n_r$. For the other two panels, we compute similar sums of probabilities for states $\ket{\overline{1_t, n_r}}$ and $\ket{\overline{2_t, n_r}}$.

\subsection{Experimental methods}

The transmon used in this experiment is geometrically Purcell protected using \cite{Patel2025WISPE}. To enhance our ability to distinguish multiple qubit states simultaneously, an RF-SQUID-based Josephson parametric amplifier is used to increase readout signal-to-noise ratio \cite{Kaufman2024SimpleAmps}. Qubit and resonator signals are generated and processed using an RFSoC ZCU216 with QICK firmware \cite{Ding2024QICK}.

Initial power calibration of the qubit drive pulse is performed by measuring the signal power just outside the input line assembly of the dilution refrigerator at room temperature using a Keysight CXA N9000B spectrum analyzer. The drive voltage is inferred from the measured power assuming the impedance is $\SI{50}{\Omega}$. To calibrate measured voltage $V_d$ to $\varepsilon_d$ (y-axis in \cref{fig:Figure_1}a), we compare the positions of the first three Rabi fringe peaks of the $\omega_{01}/3$ subharmonic resonance in experiment and simulation. A linear fit of the form $V \propto (\hbar/e)\varepsilon_d$ provides a direct conversion from drive voltage to drive amplitude.

\subsection{Floquet analysis}

For the Floquet analysis, we simplify the system by neglecting the readout resonator and focus on tracking the Floquet modes and quasienergies of the propagator $\hat U(T, 0)$ as the drive strength increases \cite{Cohen2023Reminiscence,Dumas2024}. Importantly, the propagator $\hat U(T, 0)$ accounts for the contributions of higher-order Josephson harmonics in the transmon Hamiltonian, see Supplementary Information.

In unbounded Hamiltonians such as the transmon Hamiltonian, the Floquet spectrum features avoided crossings with quasienergy gaps of arbitrarily small sizes~\cite{Hone1997Time-dependent}. These weak avoided crossings are traversed diabatically~\cite{Drese1999Floquet,Wang2025} with high probability during our fast ramp-up and ramp-down pulses and are, therefore, irrelevant to the analysis. To capture only the features relevant to the experiment, we compute the Floquet spectrum as a function of $\varepsilon_d/2\pi$ with a finite increment of $\SI{10}{MHz}$~\cite{Dumas2024}. This has the effect of skipping most avoided crossings of size $\lesssim \SI{1}{MHz}$. With the Floquet modes tracked, we compute the quasienergies and the average transmon population of each Floquet mode as a function of the drive amplitude (\cref{fig:Figure_2}a,b).

The drive strength threshold is determined using a k-means clustering algorithm applied to the $n_g$-averaged population of the transmon Floquet modes. For fixed $\omega_d$ and $\varepsilon_d$, the averaged populations are grouped into two clusters by minimizing within-cluster variance. One of these clusters corresponds to the chaotic manifold, which can be readily identified by visual inspection in~\cref{fig:Figure_2}b. We define the average population of the chaotic cluster as $M$, and use $M - 2$ as a cutoff to account for fluctuations within the cluster. We define the threshold as the lowest drive amplitude at which the average population of the ground-state Floquet mode exceeds $M-2$. Repeating this procedure for all $\omega_d$ yields the Floquet-based threshold prediction shown in~\cref{fig:Figure_1}. Applying the same method to the Floquet modes $\ket{\phi_{1_t}}$ and $\ket{\phi_{2_t}}$ gives the thresholds shown in~\cref{fig:Figure_3}.


\section*{Acknowledgements}
We thank Benjamin Groleau-Paré for valuable discussions on classical chaos in the transmon and for generating the Poincaré section plots.
This work is supported by a collaboration between the U.S. Department of Energy, Office of Science, National Quantum Information Science Research Centers, the Co-design Center for Quantum Advantage (C2QA) under Contract No.~DE-SC0012704 and the Quantum Systems Accelerator. Additional support is acknowledged from the National Agency for Research and Development (ANID) through FONDECYT Postdoctoral Grant No.~3250130, NSERC, the Ministère de l’Économie et de l’Innovation du Québec, and the Canada First Research Excellence Fund.


\bibliography{references}

\end{document}


\title{Supplementary Information: Exceeding the Parametric Drive Strength Threshold in Nonlinear Circuits}

\author{Mingkang Xia}
\altaffiliation{These authors contributed equally to this work.}
\affiliation{Department of Physics and Astronomy, University of Pittsburgh, Pittsburgh, PA 15213, USA}
\affiliation{Department of Applied Physics, Yale University, New Haven, CT 06511, USA}
\author{Crist{\'o}bal Lled{\'o}}
\altaffiliation{These authors contributed equally to this work.}
\affiliation{Institut Quantique and Département de Physique, Université de Sherbrooke, Sherbrooke J1K 2R1 QC, Canada}
\affiliation{Departamento de Física, Facultad de Ciencias Físicas y Matemáticas, Universidad de Chile, Santiago 837.0415, Chile}
\author{Matthew Capocci}
\affiliation{Department of Physics and Astronomy, Northwestern University, Evanston, Illinois 60208, USA}
\author{Jacob Repicky}
\affiliation{Department of Applied Physics, Yale University, New Haven, CT 06511, USA}
\author{Benjamin D'Anjou}
\affiliation{Institut Quantique and Département de Physique, Université de Sherbrooke, Sherbrooke J1K 2R1 QC, Canada}
\author{Ian Mondragon-Shem}
\affiliation{Department of Physics and Astronomy, Northwestern University, Evanston, Illinois 60208, USA}
\author{Ryan Kaufman}
\affiliation{Department of Physics and Astronomy, University of Pittsburgh, Pittsburgh, PA 15213, USA}
\author{Jens Koch}
\affiliation{Department of Physics and Astronomy, Northwestern University, Evanston, Illinois 60208, USA}
\author{Alexandre Blais}
\affiliation{Institut Quantique and Département de Physique, Université de Sherbrooke, Sherbrooke J1K 2R1 QC, Canada}
\affiliation{Canadian Institute for Advanced Research, Toronto, M5G 1M1 Ontario, Canada}
\author{Michael Hatridge}
\affiliation{Department of Applied Physics, Yale University, New Haven, CT 06511, USA}

\date{\today}

\maketitle


\section{Experimental Details}

\begin{figure}[t!]
    \centering
    \includegraphics[width=\columnwidth]{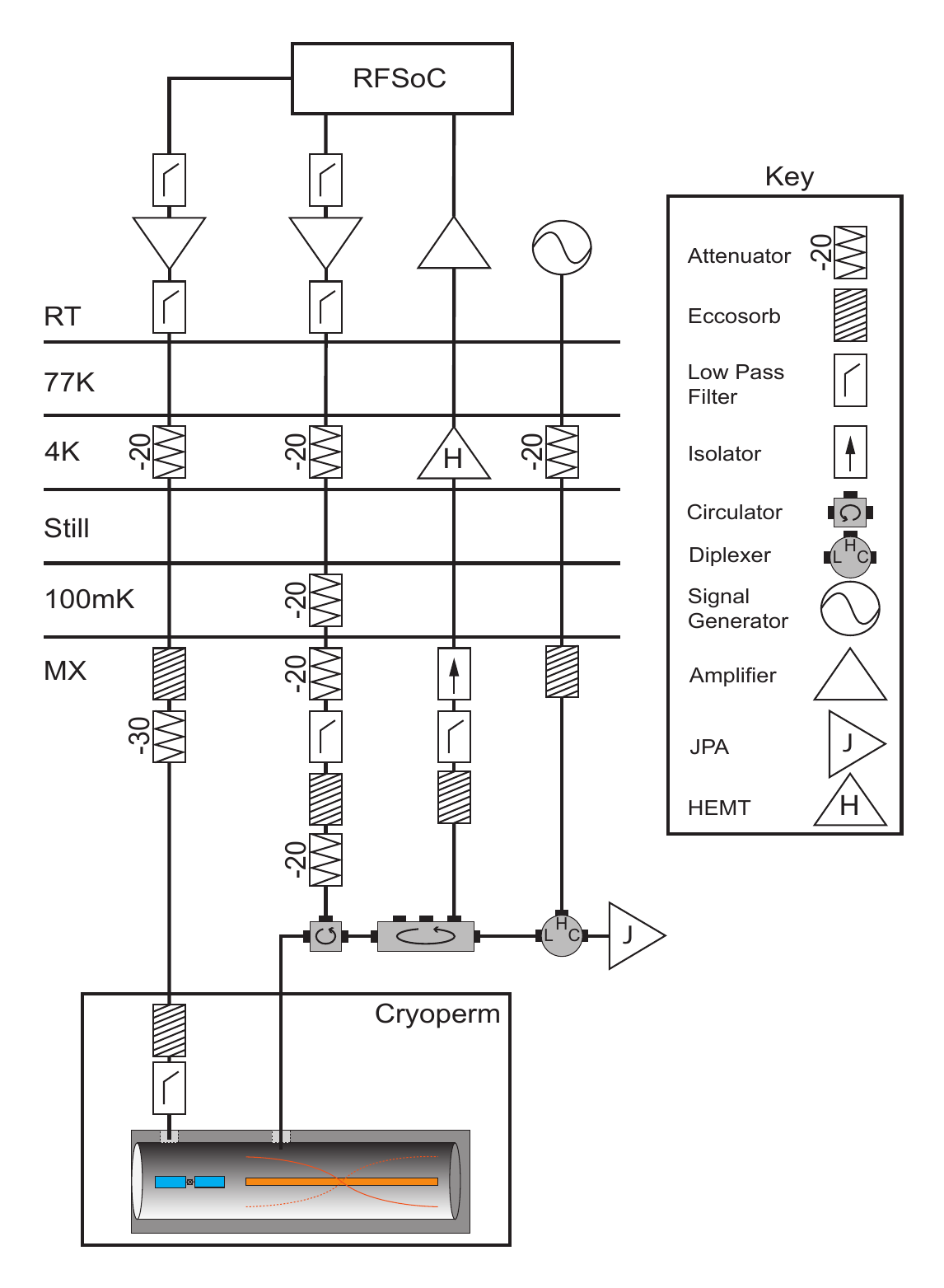}
    \caption{{\bf Diagram of the refrigerator used to perform the experiment.}
    }
    \label{fig:Fridge}
\end{figure}

A diagram of the refrigerator and sample housing used in the experiment is shown in \Cref{fig:Fridge}. Qubit and resonator signals are generated and digitized using an RFSoC ZCU216 board equipped with QICK firmware \cite{Ding2024QICK}. The qubit (blue) and $\lambda/2$ readout resonator (orange) were simultaneously fabricated out of Al using a liftoff process. Devices were deposited onto a wafer of C-plane, EFG sapphire then diced into individual chips. The selected device is held inside a \SI{4}{mm} diameter tube cut into a metal block made from Al-6061 alloy  \cite{Axline2016Architecture}. 

The readout port location was selected to achieve strong coupling to the resonator while limiting coupling to the qubit mode ($Q_{ext}=10^7-10^8$) \cite{Patel2025WISPE}. The qubit drive port is designed to be strongly coupled ($Q_{ext}=10^6$) to avoid heating the base stage of the dilution refrigerator with high power drives. We protect qubit coherence by installing a reflective low-pass filter (Mini-Circuits ZLSS252-100W-S+) as close as possible to the drive port, outside of the housing. 

\begin{figure}[t]
    \centering
    \includegraphics[width=\columnwidth]{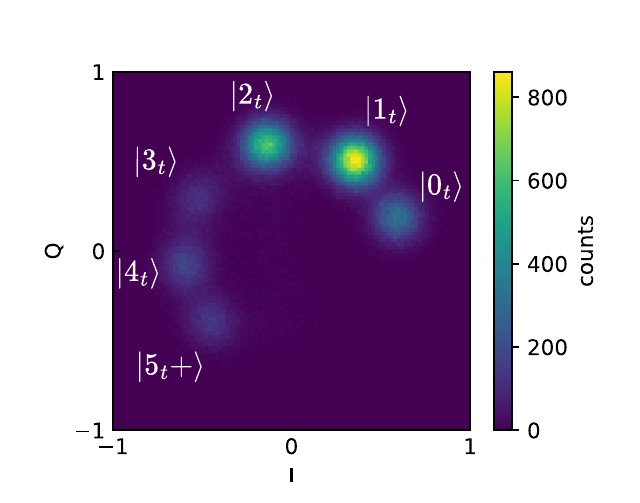}
    \caption{
    {\bf Histogram of measurements performed on the transmon that differentiate its first six states.} 
     The transmon was prepared in the $\ket{5_t}$ state by sequentially applying $\pi$-pulses corresponding to the resonant transitions. During this process, the transmon can decay back to the ground state. The measurement is performed with a $\SI{1}{\mu s}$ flat-top pulse.
    }
    \label{fig:histogram}
\end{figure}

The transmon was designed with the readout parameters ($\chi/2\pi=\SI{0.4}{MHz}$, $\kappa/2\pi=\SI{1.7}{MHz}$, $\omega_{r}/2\pi=\SI{6404.3}{MHz}$) and maintains relatively high coherence despite its strong coupling to the drive port ($T_1=\SI{32}{\mu s}$, $T_{2R}=\SI{44}{\mu s}$, $T_{2E}=\SI{62}{\mu s}$). The coupling between the qubit and readout resonator is small to better distribute the Gaussian histograms corresponding to different qubit states about the IQ plane as shown in \cref{fig:histogram}. This allows simultaneous readout of states up to $\ket{4_t}$ when using a parametric amplifier to further improve readout fidelity~\cite{Kaufman2024SimpleAmps}. The frequency of the pulse used for readout is shifted by $-2\chi$ from $\omega_{r,0_t}$ to increase sensitivity to higher excited transmon states. Under this readout condition, states $\ket{5_t}$ and above produce overlapping histograms and cannot be reliably distinguished. Therefore, the most distant blob from $\ket{0_t}$ in \cref{fig:histogram} is labeled $\ket{5_t+}$.

\section{Transmon excited to many levels}

\begin{figure*}[t]
    \centering
    \includegraphics[width=2\columnwidth]{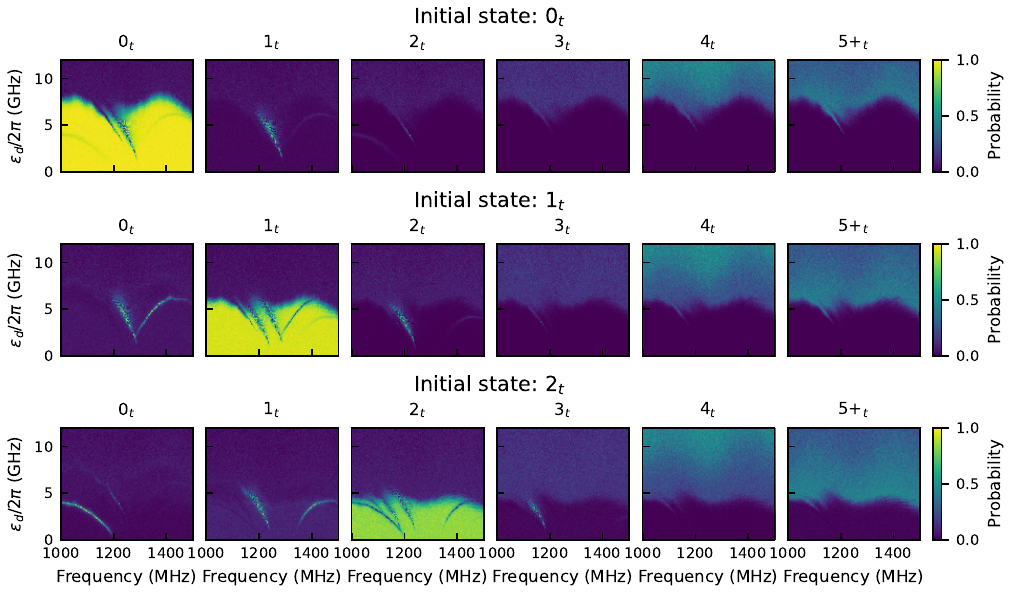}
    \caption{{\bf Effect of transmon initial state on breakdown threshold.}
    }
    \label{fig:prep_states}
\end{figure*}

 \Cref{fig:prep_states} shows the result of three experiments where the transmon is prepared in the $\ket{0_t}$, $\ket{1_t}$, or $\ket{2_t}$ states before applying the 100 ns drive. We read out the transmon and present the occupation probability of states up to $\ket{5_t+}$ for each initial state.
 nterestingly, beyond the drive amplitude threshold, the initial state has little noticeable impact on the final occupation distribution, apart from the consistently lower threshold observed for higher-energy initial states, as discussed in the main text.

\begin{figure*}[t]
    \centering
    \includegraphics[width=2\columnwidth]{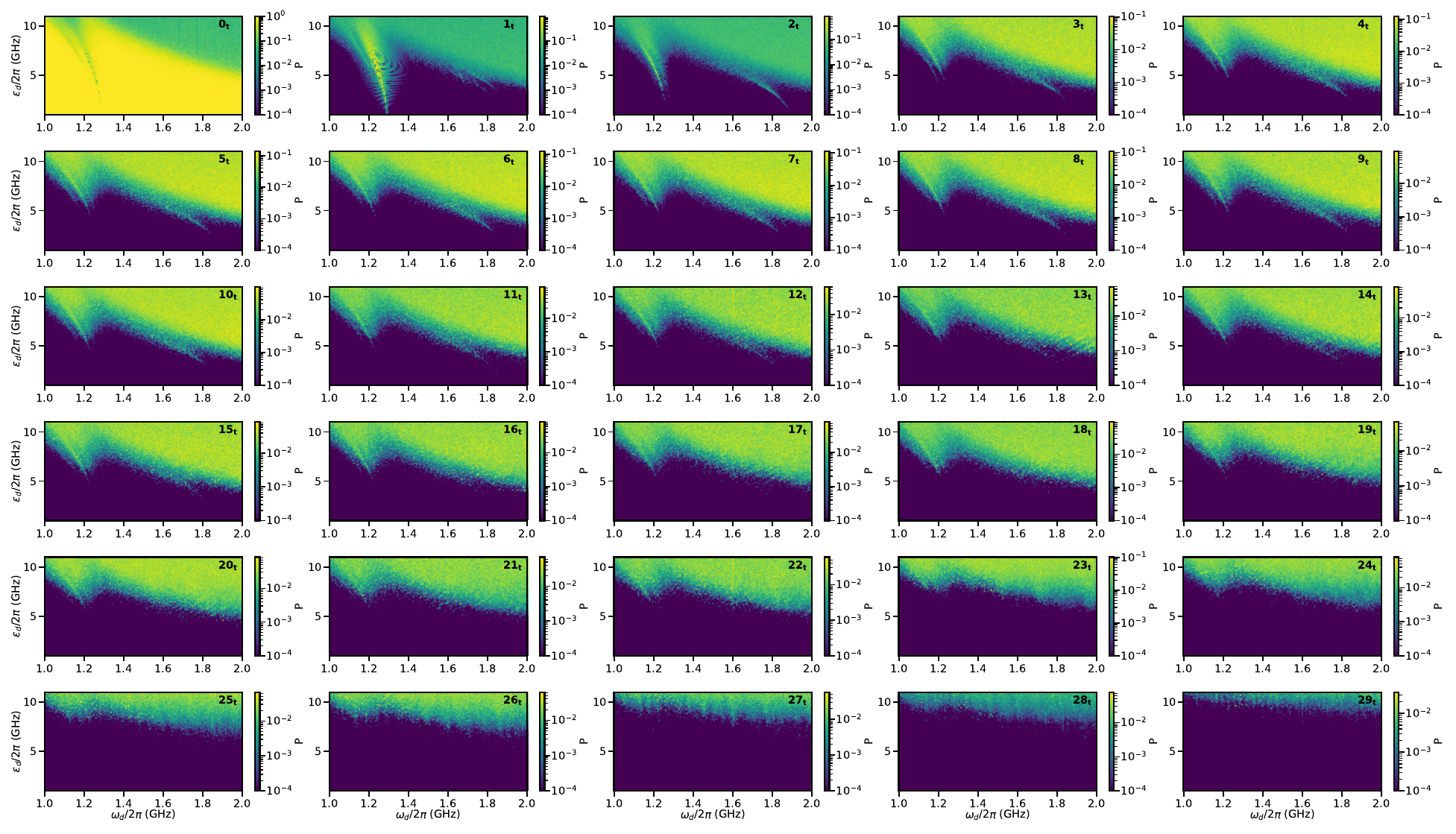}
    \caption{{\bf Transmon distribution after the pulse in simulations.}
    These are the same results as in~Fig.1b  of the main text, but here we show the probability distribution in the first thirty levels and in logarithmic scale.
    }
    \label{fig:Time_dynamics_all_levels}
\end{figure*}

To further illustrate that the transmon becomes excited into many levels, we turn to numerical simulations. \Cref{fig:Time_dynamics_all_levels} shows the results of simulations performed using the same approach as in Fig.~1b, but now displaying the occupation probabilities of the transmon's first thirty states after the 100 ns-long pulse. At sufficiently high drive amplitude, the distribution is observed to spread among all these thirty states.

\section{Spectroscopy measurements and model fitting}

The qubit transition frequency $\omega_{i_t, i_t+1}$ is measured by first preparing the transmon in the $\ket{i_t}$ state using a sequence of resonant pulses. A weak probe tone is then applied, sweeping across frequency to identify the transition to the $\ket{i_t+1}$ state. The $\omega_{01}$ transition is characterized more precisely via a Ramsey experiment, where the observed oscillations as a function of time reveal the detuning between the drive and the qubit. The resonator frequencies $\omega_{r,0}$ and $\omega_{r,1}$, corresponding to the transmon being in states $\ket{0_t}$ and $\ket{1_t}$, respectively, are determined by preparing the transmon in the respective state, applying a weak probe tone to the resonator, and fitting the decay trace to extract $\chi$. The frequencies $\omega_{r,0}$ and $\omega_{r,1}$ are inferred from those measurements.

\begin{figure}[t]
    \centering
    \includegraphics{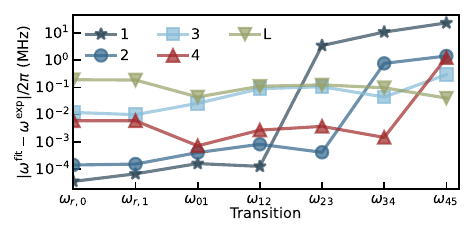}
    \caption{
    {\bf Comparison between measured transition frequencies and several fitted theoretical models.}
    The first four transmon transition frequencies and two conditional resonator shifts are fitted using several transmon models. A standard transmon model with a single Josephson harmonic (N = 1, blue stars) fails to capture the measured transitions. Including a series stray inductance (green triangles) remains insufficient. In contrast, treating the energies of higher-order Josephson harmonics as independent fitting parameters (N = 2: blue circles; N = 3: blue squares; N = 4: red triangles) yields progressively better agreement. With four harmonics, all measured transitions are reproduced within or near their experimental uncertainties, as discussed in the text.
    }
    \label{fig:model_fitting}
\end{figure}

The two conditional resonator frequencies, together with the first four transmon transition frequencies, are then fitted to a range of theoretical transmon qubit models, as shown in \cref{fig:model_fitting}. The standard transmon model truncated to a single Josephson harmonic (N=1, blue stars) fails to reproduce the experimentally measured transition frequencies. 

We also consider a model that includes a linear stray inductance $L$ in series with the Josephson junction. Assuming that the Junction plasma frequency is much larger than the transmon frequency, we neglect the contribution from the small Junction capacitance. The resulting Lagrangian reads
\begin{equation}
\mathcal{L}(\varphi, \varphi_L) = \frac{C}{2} \left(\frac{\Phi_0}{2\pi}\right)^2 \dot{\varphi}^2 + E_J \sin(\varphi - \varphi_L) - \frac{E_L}{2} \varphi_L^2,
\end{equation}
where $C$ is the shunt capacitance, $E_L = \Phi_0^2/[L(2\pi)^2]$ the inductive energy, $\varphi$ is the reduced total flux across the junction and inductor, and $\varphi_L$ is the reduced flux across the inductor alone. Following the approach of \textcite{Willsch2024_HigherTransmonHarmonics}, we treat the Lagrangian perturbatively in the small \textit{screening parameter} $E_J/E_L \ll 1$.
Applying the Euler–Lagrange equation to $\varphi_L$ yields the current conservation condition
\begin{equation}
E_L \varphi_L = E_J \sin(\varphi - \varphi_L),
\end{equation}
which admits a unique solution when $E_J/E_L<1$. We solve this equation perturbatively in powers of $E_J/E_L$ to obtain an approximate expression for $\varphi_L(\varphi)$ and substitute it back into the Lagrangian. This yields an effective single-variable Lagrangian $\mathcal L(\varphi)$, whose potential energy we expand in powers of $E_J/E_L$ up to the third order. Since $\varphi_L$ is a periodic function of $\varphi$, the resulting potential includes contributions from four effective Josephson harmonics via the Jacobi-Anger expansion. Importantly, the corresponding Josephson energies are not treated as independent fitting parameters, but are instead fixed functions $E_J$ and $E_L$ whose values we fit. As shown in \cref{fig:model_fitting} (L, green triangles), this model does not adequately reproduce the measured transition frequencies.

In contrast, a model in which the amplitudes of the higher-order Josephson harmonics (N = 2, blue circles; N = 3, blue squares; N = 4, red triangles) are treated as independent fitting parameters yields progressively better agreement with the experiment. Using four harmonics, the model reproduces all transmon transitions to within or near their respective experimental uncertainties, which range from $\SI{0.4}{kHz}$ for $\omega_{01}/2\pi$ to $\SI{3}{MHz}$ for $\omega_{45}/2\pi$. The small deviation in $\omega_{01}$ (about $\SI{0.7}{kHz}$) slightly exceeds the nominal uncertainty, while the larger uncertainty in $\omega_{45}$ arises from preparation errors and gate charge fluctuations. This transition is, therefore, reproduced with correspondingly less precision.

This modeling approach is consistent with recent findings from several groups, who have reported similar improvements in fitting transmon spectra by including higher-order Josephson harmonics in devices fabricated using different methods~\cite{Willsch2024_HigherTransmonHarmonics,Fechant2025,Wang2025}.

The simulations presented here and in the main text use $N=4$ harmonics with the following fitted parameters: $\omega_r/2\pi = \SI{6.4043}{GHz}$, $g/2\pi = \SI{60}{MHz}$, $E_C/2\pi = \SI{149.6}{MHz}$, $E_{J,1}/2\pi = \SI{14.0286}{GHz}$, and $(E_{J,2}, E_{J,3}, E_{J,4})/2\pi = (-142.5,\ 8.4,\ -2.3)~\text{MHz}$.

\begin{figure}[t]
    \centering
    \includegraphics{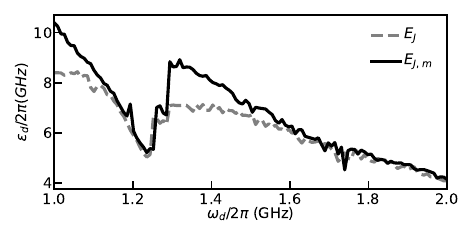}
    \caption{{\bf Quantitative comparison between single- and multi-harmonic transmon models.}
    Critical drive amplitude extracted from Floquet analysis using either a standard transmon model with a single Josephson harmonic ($E_J$) or a model with four independent Josephson harmonics ($E_{J,m}$) as used throughout our work.}
    \label{fig:Floquet_prediction_single-vs-multiple-harmonics}
\end{figure}

To quantify the impact of including higher-order Josephson harmonics, we compare in \cref{fig:Floquet_prediction_single-vs-multiple-harmonics} the critical drive amplitude extracted from Floquet analysis using multiple harmonics (black full line, also shown in Fig.~1 of the main text) to that obtained with a single-harmonic model (dashed gray line). The threshold can differ by as much as $\SI{2}{GHz}$ at low drive frequencies, where higher transmon levels contribute significantly to the strong hybridization of Floquet modes. These higher levels differ substantially between the two models, resulting in marked discrepancies in the predicted threshold. See also \textcite{Fechant2025} for a discussion of how higher harmonics influence qubit measurement-induced state transitions.

\section{Hybridization at positive detuning}

\begin{figure*}[t]
    \centering
    \includegraphics{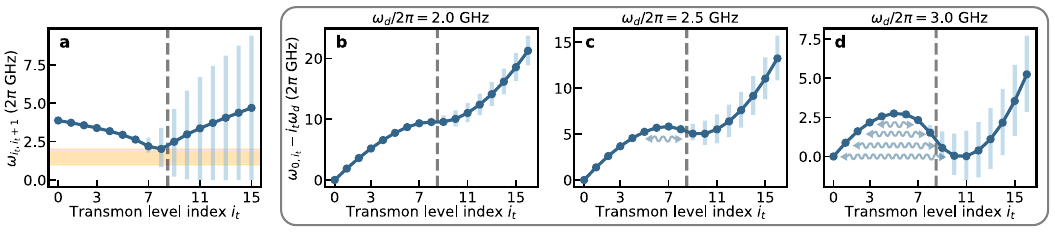}
    \caption{
    {\bf Absence of direct $(n:n)$ resonances in the transmon spectrum for large detunings.}
    {\bf a}, Consecutive bare transition frequencies of the transmon at $n_g=0.25$. Vertical light-blue solid lines represent the charge dispersion at level $i_t$. The area shaded orange represents the drive frequency window $\omega_d/2\pi\in [1, 2]\, \text{GHz}$ of this work and illustrates that a direct one-photon resonance is only possible at the top of the well (vertical dashed line) or for higher levels, for some realizations of the gate charge.
    {\bf c-d}, Detuning between bare transmon transitions $\omega_{0_t, i_t}$ at $n_g=0.25$ and $i_t$ drive photons at $\omega_d$, for three different drive frequencies. The vertical blue lines represent charge dispersion. The arrows represent $(n:n)$ resonances, namely direct multiphoton processes where the transmon absorbs $n$ photons from the drive to climb $n$ levels. These resonances, which lead to the bunching of levels inside the cosine potential and can precipitate ionization for the computational states, appear typically in the detuning regime of measurement-induced ionization~\cite{Khezri2023Measurement,Dumas2024}.
    For the large detuning regime $\omega_d/2\pi < \SI{2}{GHz}$ of this work, there are no such direct $(n:n)$ resonances, except for a one-photon resonance at the top of the transmon cosine potential, indicated by the vertical dashed line, or above.
    }
    \label{fig:hybridization}
\end{figure*}

To understand the frequency dependence of the ionization threshold observed in the experiment and simulations, we examine the structure of the transmon spectrum and contrast it with the situation where unwanted transitions are induced by a readout drive, and thus typically with much smaller detuning than used here~\cite{Khezri2023Measurement,Dumas2024}.
\Cref{fig:hybridization}a shows the consecutive bare transition frequencies $\omega_{i_t, i_t+1}$ of our transmon at a fixed gate charge $n_g = 0.25$. The vertical light-blue bars represent the total charge dispersion for each transition. The width of these bars increases with level index, reflecting the enhanced charge sensitivity of eigenstates near and above the top of the cosine potential, marked by the vertical dashed line.

The orange-shaded area indicates the drive frequency range considered in this work, $\omega_d/2\pi \in [1, 2]~\text{GHz}$. Direct one-photon transitions between low-lying transmon states are energetically forbidden due to the large transmon-drive detuning $\omega_{i_t, i_t+1} - \omega_d$. A one-photon resonance is only possible for transitions at or above the top of the cosine potential. For those higher states, hybridization with the drive becomes significant and nonperturbative even at modest amplitudes, when $\varepsilon_d \sim\omega_{i_t, i_t+1} - \omega_d$, as discussed in the main text.

Other types of resonances can also play a role. Inspired by Ref.~\cite{Khezri2023Measurement}, \cref{fig:hybridization}c–d display the detuning $\omega_{0_t, i_t} - i_t \omega_d$ as a function of $i_t$ for three representative drive frequencies. This reveals potential $(n:n)$ resonances, where the absorption of $n$ photons leads to excitation by $n$ levels. Arrows in the figure highlight such processes. These resonances can lead to level bunching deep within the cosine potential and precipitate ionization from the computational subspace~\cite{Dumas2024}. In regimes where these resonances are present, the ionization threshold exhibits an oscillatory dependence on drive frequency---each dip in the threshold is associated with a specific $(n:n)$ resonance. This is the typical behavior in transmon readout settings at positive qubit-resonator detuning~\cite{Khezri2023Measurement}, where reduced detuning enhances the dispersive shift $\chi$.

In contrast, for the large positive qubit-drive detuning used here, no $(n:n)$ multiphoton resonances are encountered within the operating frequency window (\cref{fig:hybridization}b). This is consistent with the monotonic, smooth frequency dependence of the ionization threshold observed in Fig.~1 of the main text and confirms that in this regime, no single resonance dominates the ionization.

\section{Poincaré section}
\label{sm:sec:poincare}

The classical limit of the transmon corresponds to the rotor or rigid pendulum~\cite{Koch2007ChargeInsensitive}, obtained by taking $E_{J,m} \to \infty$ and $E_C \to 0$ while keeping their product constant. Since our system includes multiple Josephson harmonics, we first define $E_{J,m} = c_m E_J$, where the coefficients $c_m = E_{J,m}/E_J$ remain fixed. We choose to fix the product of Josephson and charging energies by the plasma frequency $\omega_p = \sqrt{8 E_C E_J} = \sqrt{8 E_C E_{J,1}} = 2\pi \times \SI{652.1}{MHz}$.

We introduce the rescaled charge operator $\tilde{n}_t = z \hat{n}_t$, where $z = \sqrt{8E_C / E_J}$ is the transmon impedance. Using the transmon Hamiltonian,
\begin{equation}
\begin{split}
\hat{H}(t) = & 4E_C (\hat{n}_t - n_g)^2 - \sum_m E_{J,m} \cos(m \hat{\varphi}_t) \\
&+ \varepsilon_d \sin(\omega_d t) \hat{n}_t,
\end{split}
\end{equation}
we derive the Heisenberg equations for $\tilde{n}_t$ and $\tilde{\varphi}_t$, where $\tilde{\varphi}_t = \hat{\varphi}_t$ is left unscaled. Taking the classical limit yields
\begin{equation}
\begin{split}
\partial_t \tilde{n}_t &= -\omega_p \sum_m m c_m \sin(m \tilde{\varphi}_t), 
\\
\partial_t \tilde{\varphi}_t &= \omega_p \tilde{n}_t + \varepsilon_d \sin(\omega_d t).
\end{split}
\end{equation}
In this limit, the commutator $[\tilde{n}_t, \tilde{\varphi}_t] = -iz$ between the rescaled operators---and any other nested commutator of them---tend to zero for any finite drive amplitude, implying that all operator expressions can be replaced by their classical counterparts: $\tilde{n}_t$ and $\tilde{\varphi}_t$ can be treated as classical variables.

\begin{figure*}
    \centering
    \includegraphics[width=1\linewidth]{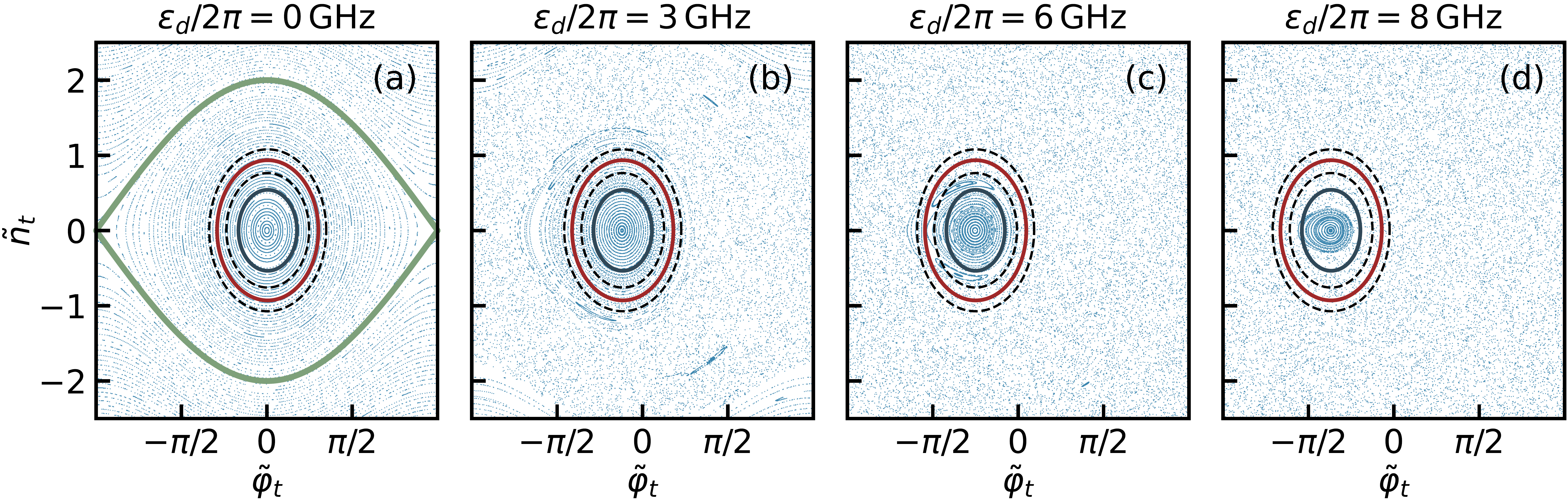}
    \caption{
    {\bf Growth of chaos in the Poincaré sections of the classical analog of the transmon.
    }
    The phase space trajectories $\{\tilde{\varphi}_t(t), \tilde{n}_t(t)\}$ are plotted stroboscopically at integer multiples of the drive period for many different initial conditions and four representative values of the drive amplitude. The drive frequency is $\omega_d/2\pi=\SI{1.75}{GHz}$. Chaos emerges around the separatrix, the green orbit in panel (a), for an arbitrarily small drive amplitude. The chaos region grows as $\varepsilon_d$ increases, eventually swallowing the quantized Bohr-Sommerfeld orbits corresponding to the Floquet modes $1_t$ (red) and $0_t$ (blue).
    }
    \label{fig:Poincare}
\end{figure*}

To visualize the resulting dynamics, we use Poincaré sections which show the phase-space trajectories $\{\tilde{\varphi}_t(t), \tilde{n}_t(t)\}$ stroboscopically at integer multiples of the drive period $T = 2\pi / \omega_d$ for many different initial conditions. \Cref{fig:Poincare} shows representative Poincaré sections for $\omega_d/2\pi =\SI{1.75}{GHz}$ and increasing drive amplitudes.

In panel (a), for $\varepsilon_d = 0$, two types of motion are observed: (i) bounded, closed orbits around the origin, and (ii) unbounded, rotating orbits corresponding to full pendulum swings. These are separated by a special trajectory called the separatrix (green), which is highly sensitive to perturbations. A chaotic layer emerges around the separatrix when a small drive is applied. As the drive amplitude increases, this chaotic region grows, as seen in panels (b)–(d).

Bohr-Sommerfeld quantization provides a semiclassical link between this classical picture and quantum dynamics. Quantized periodic orbits in the Poincaré section correspond to quantum Floquet modes. Each such orbit is associated with an area of $2\pi \hbar_\text{eff} = 2\pi z$ in phase space, representing the quantum uncertainty around the trajectory. The quantization condition requires the $i_t$-th orbit to enclose an area $2\pi z (i_t + 1/2)$~\cite{Messiah2020Quantum}.

Following Refs~\cite{Cohen2023Reminiscence,Dumas2024}, in \cref{fig:Poincare}, we show quantized classical orbits for the ground ($0_t$, red) and first excited ($1_t$, blue) Floquet states. The dashed black contours represent the required phase-space area for quantum fluctuations. Both orbits fit comfortably within the central regular region in panels (a) and (b). However, in panel (c), the chaotic layer occupies so much area that the $1_t$ orbit can no longer be defined. In panel (d), even the $0_t$ orbit is lost. This swallowing of the quantized orbits in the semiclassical picture is a good predictor of ionization of the computational states in the quantum system~\cite{Dumas2024}.

\bibliography{SM_references}